\documentclass[aps, showpacs, showkeys,nofootinbib,floatfix]{revtex4}

\usepackage{amssymb}
\usepackage{amsmath}
\usepackage{graphicx}
\usepackage{hyperref}
\usepackage{graphicx}

\begin{document}

\title{Thermodynamics of Apparent Horizon and Friedmann Equations in Big Bounce Universe}
\author{Molin Liu$^{1}$}
\thanks{Corresponding author\\E-mail address: mlliu@xynu.edu.cn}
\author{Yuling Yang$^{1}$}
\author{Jianbo Lv$^{2}$}
\author{Lixin Xu$^{3}$}
\affiliation{$^{1}$College of Physics and Electronic Engineering,
Xinyang Normal University, Xinyang, 464000, P. R. China\\
$^{2}$Department of Physics, Liaoning Normal University, Dalian, 116029, P. R. China\\
$^{3}$School of Physics and Optoelectronic Technology, Dalian University of
Technology, Dalian, 116024, P. R. China}


\begin{abstract}
In this paper, we study a big bounce universe typified by a non-singular big bounce, as opposed to a singular big bang. This cosmological model can describe radiation dominated early universe and matter dominated late universe in FRW model. The connections between thermodynamics and gravity are observed here. In the early stage of both cold and hot universes, we find there is only one geometry containing a 4D de Sitter universe with a general state parameter. We also find the form of the apparent horizon in the early universe strongly depends on the extra dimension, which suggests that the influence of the extra dimension could in principle be found in the early universe. Moreover, we show that in the late stages of both cold and hot universes, the moment when the apparent horizon begins to bounce keeps essentially in step with the behavior of the cosmological scalar factor.
\end{abstract}



\maketitle

\section{Introduction}
Recently, the holographic principle points out that a N dimensional gravity theory could be equivalent to a (N $-$ 1) dimensional theory without gravity, such as the AdS/CFT correspondence \cite{Maldacena}. The holographic principle implies that some deep connections might exist between thermodynamics and gravity. One pioneer work of this subject was presented by Jacobson in Ref.\cite{Jacobson}, in which the Einstein field equation was obtained by using the Clausius relation $\delta Q = T \delta S$ where $\delta Q$ is the energy flux, $T$ is Unruh temperature and $S$ is the entropy of thermodynamic system. Then, Brustein and Hadad proved that the motion equation of generalized gravity theory is equivalent to thermodynamic Clausius relation through a more general definition of the Noether charge entropy \cite{Brustein}. Soon after that, this method is generalized to diffeomorphism invariant theories of gravity by Padmanabhan in Ref.\cite{Padmanabhan}, and Parikh and Sarkar in Ref.\cite{Parikh}. On the other hand, the inverse view was put forward to build the connection in which the Einstein equation could be treated as a thermodynamic identity. Padmanabhan first noticed that the gravitational field equation of the spherically symmetric spacetime can be rewritten in a form of the ordinary first law of thermodynamics at the black hole horizon \cite{Padmanabhan2}. Then, Cai and Kim employed different kinds of horizons with given entropy to obtain the Friedmann equations from the first law of thermodynamics in the different corresponding gravity theories including $(n + 1)$ dimensional FRW universe with general spatial curvature, the higher derivative Gauss-Bonnet gravity and Lovelock gravity \cite{cairgKimsp}. Under FRW cosmological setup, the relationships between thermodynamics and gravitational field equation have been intensively investigated by various gravity models, such as $f(R)$ gravity \cite{frgravity}, scalar tensor gravity \cite{scalartensor}, braneworld cosmology \cite{brancase}. About the reviews of this aspect, one can refer to Refs.\cite{Padmanabhan3,Padmanabhan4}.

In cosmology, the question of whether time exists before the so-called ``big bang" has always been an interesting topic \cite{Steinhardt,Tolman,Khoury1,Khoury2,LiuWesson}. Tolman proposed an oscillating cosmological model in the context of general relativity and pointed out that the universe had to face the cosmological singularity and enormous inhomogeneities in every evolutional cycle \cite{Tolman}. Early in this century, under the brane world scenario the so-called ``ekpyrotic" cosmological model was presented by Khoury et.al. in Refs.\cite{Khoury1,Khoury2}. Then, Steinhardt and Turok presented a cyclic model where universe is dynamical from the big bang to the big crunch in every cycle \cite{Steinhardt}. Most cosmological problems such as the horizon problem and flatness problem could be solved well without inflation in the above various cosmological models \cite{Steinhardt,Tolman,Khoury1,Khoury2}.

Based on the consideration about the problems of the cosmological constant and the singular big bang, Liu and Wesson proposed a big bounce cosmological model with a variable cosmological ``constant", in which there is only one transition from contraction to expansion \cite{LiuWesson}. Before the bounce, it contracts from an empty de Sitter vacuum. This cosmological model was derived from a class of exact solutions of the 5D field equations obtained by Liu and Mashhoon in Ref.\cite{LiuMashhoon}. This type of big bounce universe satisfies the 5D field equations $R_{AB} = 0$. The content of this theory is mathematically rich because two extra functions $\eta$ and $\zeta$ are included. The most interesting is that the big bang appeared in the standard 4D cosmological model is replaced by a big bounce in this model where at the bouncing point the universe arrives at a finite minimum scale. After the contraction of the bounce, our universe starts expanding immediately. In usual 4D bounce universe under the context of standard general relativity, there exists the so-called inhomogeneity problem that the enormous inhomogeneities are unavoidable in the collapsing phase \cite{Robertson}. However, in this 5D bounce universe the spacetime is not symmetric either before or after the bounce point. The contraction is dominated by the new matter created because the universe could be empty at the beginning of contraction. Hence, this situation could avoid the inhomogeneity problem appeared in the former 4D big bounce universe. It should be noticed that the function adopted as $\mu (t) = t^n$ is based on the fact that in the spatially flat FRW universe the time evolution of the scale factor has the form of power law. Considering two different phases in early and late universe, the radiation dominated and matter dominated standard FRW universes are obtained respectively. By employing the first law of thermodynamics and the fluid's continuity equation to the apparent horizon, we could derive the Friedmann equations in the 5D big bounce universe. Motivated by the recent astronomical observations which favor a flat universe, we also study its universal spatially flat model and its two extreme cases: the cold 3D flat case and the hot 3D flat case. Based on the above works, the motivation of this paper is trying to study the thermodynamics of the apparent horizon and Friedmann equations and to obtain the connections between the thermodynamics and gravity for the big bounce universe.

This paper is organized as follows: In Section II, the 5D big bounce cosmological model is presented and the general Friedmann equations are derived through the thermodynamics of the apparent horizon. In Section III, thermodynamics of the apparent horizon is discussed in the general spatially flat model. In the first and second subsections, the thermodynamics of the apparent horizon are discussed in the late cold/hot 3D flat model respectively. In the third subsection, the early stage of cold/hot 3D flat universes are investigated. Section IV is the conclusion. We adopt the signature ($+$, $-$, $-$, $-$, $-$) and set $\hbar$ and $c$ equal to unity. Greek indices $\mu$, $\nu$, $\cdots$ will be taken to run over 0, 1, 2, 3 as usual, while capital indices $A$, $B$, $C$, $\cdots$ run over all five coordinates (0, 1, 2, 3, 5).
\section{the big bounce solutions and their thermodynamics of the apparent horizon}
A class of cosmological solutions with extra dimensions were presented firstly by Liu and Mashhoon \cite{LiuMashhoon} and were extended to the version containing a big bounce by Liu and Wesson \cite{LiuWesson}. The big bounce cosmological solutions read as follows,
\begin{equation}\label{15dmetric}
d S^2 = \zeta^2(t,y) d t^2 - \eta^2(t,y)\left[\frac{dr^2}{1-kr^2} + r^2 \left(d\theta^2 + \sin^2\theta d \phi^2 \right)\right] - d y^2,
\end{equation}
which satisfies the 5D field equations $R_{AB} = 0$. Here $k$ is the 3D curvature index ($k = \pm 1$ or $0$) and the scale factors $\zeta$ and $\eta$ are listed as
\begin{eqnarray}
  \eta^2 &=& \left(\mu^2 + k\right) y^2 + 2\nu y + \frac{\nu^2 + K}{\mu^2 + k},\label{addeta}\\
  \zeta &=& \frac{1}{\mu} \frac{\partial \eta}{\partial t} \equiv \frac{\dot{\eta}}{\mu},\label{addzeta}
\end{eqnarray}
where $\mu = \mu(t)$ and $\nu = \nu(t)$ are two arbitrary functions varying with time. $K$ is a 5D curvature constant related to the Riemann-Christoffel tensor via
\begin{equation}\label{1Kretschmann}
R_{ABCD}R^{ABCD} = \frac{72K^2}{\eta^8},
\end{equation}
which agrees with the case of canonical coordinates models \cite{LiuMashhoon}. In this 5D universe, there is a freedom to fix $\mu = \mu(t)$ or $\nu = \nu(t)$ without changing the form of solutions (\ref{15dmetric}) because $\zeta (t,y) d t$ is invariant under arbitrary transformation.

The 4D component of above 5D metric (\ref{15dmetric}) could be written in the Robertson-Walker form under the standard FRW model as
\begin{equation}\label{144lineelement}
d s^2 = \zeta^2(t,y) d t^2 - \eta^2(t,y)\left[\frac{dr^2}{1-kr^2} + r^2 \left(d\theta^2 + \sin^2\theta d \phi^2 \right)\right].
\end{equation}
The non-vanishing components of the 4D Ricci tensor are given by
\begin{eqnarray}
  ^{(4)} R_0^0 &=& -\frac{3}{\zeta^2}\left(\frac{\ddot{\eta}}{\eta} - \frac{\dot{\eta}\dot{\zeta}}{\eta\zeta}\right), \\
  ^{(4)} R_1^1 &=& ^{(4)} R_2^2 = ^{(4)} R_3^3 = -\frac{1}{\zeta^2} \left[\frac{\ddot{\eta}}{\eta} + \frac{\dot{\eta}}{{\eta}}\left(\frac{2\dot{\eta}}{\eta} - \frac{\dot{\zeta}}{\zeta}\right) + 2 k \frac{\zeta^2}{\eta^2} \right].
\end{eqnarray}
So the 4D Ricci scalar is
\begin{equation}\label{14dRicciscalar}
^{(4)}R = - 6 \left(\frac{\mu\dot{\mu}}{\eta\dot{\eta}} + \frac{\mu^2 + k}{\eta^2}\right).
\end{equation}
Hence, according to the 4D Einstein tensor
\begin{equation}\label{1dGalphabeta}
^{(4)}G_{\beta}^{\alpha} \equiv \ ^{(4)}R_{\beta}^{\alpha} - \delta_{\beta}^{\alpha}\ ^{(4)}R/2,
\end{equation}
the non-vanishing components are obtained as
\begin{eqnarray}
  ^{(4)}G_{0}^{0} &=& \frac{3\left(\mu^2 + k\right)}{\eta^2}, \\
  ^{(4)}G_{1}^{1} &=& ^{(4)}G_{2}^{2} = ^{(4)}G_{3}^{3} = \frac{2\mu}{\eta} \frac{\dot{\mu}}{\dot{\eta}} + \frac{\mu^2 + k}{\eta^2}.
\end{eqnarray}

According to the proper time defined by $d \tau = \zeta(t, y) d t$, we can get the Hubble parameter written as,
\begin{equation}\label{1hubble}
H(t, y) \equiv \frac{1}{\zeta} \frac{\dot{\eta}}{\eta} = \frac{\mu}{\eta}.
\end{equation}

Then in this 5D big bounce universe model, we can get the 4D Friedmann equations shown by
\begin{eqnarray}
\label{14d5dfrieq1}  \frac{1}{\zeta} \dot{H} - \frac{k}{\eta^2} &=& -4\pi G \left(p + \rho\right),\\
\label{14d5dfrieq2}   H^2 + \frac{k}{\eta^2} &=& \frac{8\pi}{3} G\rho,
\end{eqnarray}
where the energy-momentum tensor of a perfect fluid with density $\rho$ and
pressure $p$ has the following form
\begin{equation}\label{1energymomentum}
    T^{\alpha\beta} = (p + \rho) u^{\alpha}u^{\beta} - p g^{\alpha\beta}
\end{equation}
and the 4 velocity is $u^{\alpha} \equiv d x^{\alpha}/d s$.
Interestingly, Eqs.(\ref{14d5dfrieq1}) and (\ref{14d5dfrieq2}) reduce to the usual Friedmann equations in GR gravity under the limits of $\zeta \rightarrow 1$ and $\eta \rightarrow R$,
\begin{eqnarray}
\label{1GRfrieq1}  \dot{H} - \frac{k}{R^2} &=& -4\pi G \left(p + \rho\right),\\
\label{1GRfrieq2}  H^2 + \frac{k}{R^2} &=& \frac{8\pi}{3} G\rho.
\end{eqnarray}

In order to study the connections between thermodynamics and gravity, we use a simple choice of the boundary surface based on the concept of cosmological apparent horizon which was proposed by Bak and Rey \cite{Bak}. The apparent horizon is the boundary hypersurface of an anti-trapped region and has a topology of $S^2$. It turns out that there is natural gravitational entropy associated with the apparent horizon \cite{Bak}. The 4D metric on the hypersurface of this 5D big bounce universe can be rewritten more explicitly as,
\begin{equation}\label{14dremetr}
d s^2 = h_{ab} d x^a d x^b + \widetilde{r}^2 d \Omega^2,
\end{equation}
where $d \Omega^2 = d \theta^2 +\sin^2 \theta d \phi^2$, $\widetilde{r} = \eta r$, $x^0 = t$, $x^1 = r$, $h_{ab} = \text{diag} (\zeta^2, -\eta^2/(1 - k r^2))$. Based on the above explicit spherical symmetry metric (\ref{14dremetr}), the dynamical apparent horizon is determined by the relation
\begin{equation}\label{addapparenthorizon}
h^{ab}\partial_a \widetilde{r}\partial_b \widetilde{r} = 0,
\end{equation}
which implies that the vector $\nabla \widetilde{r}$ is null on the surface of apparent horizon. The radius of apparent horizon $\widetilde{r}_A$ could be given below by the explicit evaluation of the condition,
\begin{equation}\label{1apphor1}
    \widetilde{r}_A = \frac{1}{\sqrt{H^2 + k/\eta^2}},
\end{equation}
where the apparent horizon is established on the trapped region \cite{Wald,Hawking}.

Once the geometry of spacetime $(M, g_{ab})$ is determined by the Einstein equation, the horizon area and surface gravity are purely geometric quantities determined by the spacetime geometry and the formulae of black hole entropy and temperature have a certain universality with horizon. Like in black hole physics, the associated temperature with the apparent horizon is assumed to be proportional to their surface gravity at the apparent horizon, i.e. $T = \kappa /2\pi$, where the surface gravity $\kappa$ is defined by \cite{Bak}
\begin{equation}\label{addsurface}
\kappa = \frac{1}{2\sqrt{-h}}\partial_{\mu}(\sqrt{-h} h^{\mu\nu}\partial_{\nu}\widetilde{r}).
\end{equation}
Because the structure of spacetime $(M, g_{ab})$ is spherically symmetric it is reasonable to assume that the entropy associated with the apparent horizon is one quarter of the area of the apparent horizon, i.e. $S = A_0/4G$ where $A_0 = 4 \pi \widetilde{r}_A^2$ is the area of apparent horizon.

When adopting the perfect fluid's energy-momentum tensor $T_{\mu\nu}$, the energy-supply vector is given by
\begin{equation}\label{1energysupply}
\Psi_a = T_a^b \partial_b \widetilde{r} + W \partial_a \widetilde{r},
\end{equation}
where $W$ is the work density defined by $W = -\frac{1}{2} T^{ab}h_{ab}$. By using above energy-supply vector, the amount of energy crossing the apparent horizon is given by
\begin{equation}\label{amountenergy}
    - dE = A_0 (\rho + p) \zeta H \widetilde{r}_A  dt.
\end{equation}
Based on the apparent horizon's temperature $T$, the entropy $S$ and the crossing amount energy $dE$, the first law of thermodynamics, $-dE = T d S$, can give the spatial-spatial component of the Friedmann equations shown by
\begin{equation}\label{FReq1}
    4\pi G (\rho + p) = -\left(\frac{1}{\zeta} \dot{H} - \frac{k}{\eta^2}\right),
\end{equation}
which describes the 4D FRW universe with spatial curvature $k$ in this big bounce universe.
Then based on the above Friedmann equation (\ref{FReq1}), the perfect fluid's continuity equation is shown by
\begin{equation}\label{perconeq}
    \dot{\rho} + 3 \zeta H (\rho + p) = 0,
\end{equation}
which can help us get another temporal-temporal component of Friedmann equations as
\begin{equation}\label{anotherFrea}
\frac{8\pi G}{3} \rho = H^2 + \frac{k}{\eta^2}.
\end{equation}
Hence, the Friedmann equations (\ref{FReq1}) and (\ref{anotherFrea}) for the 4D FRW universe in this 5D big bounce universe are obtained by the first law of thermodynamics and the perfect fluid's continuity equation near the apparent horizon $\widetilde{r}_A $. The connections between the gravity of 5D big bounce universe and the thermodynamics of its apparent horizon are obtained. On each hypersurface the first law of thermodynamics of apparent horizon is successfully corresponding to the spatial component of Friedmann equations of gravity. The fluid's continuity equation is corresponding to the temporal component of Friedmann equations of gravity.

In this paper, the cosmological constant $\Lambda$ could be obtained through the following scaling, namely $\rho \rightarrow \rho + \Lambda$ and $p \rightarrow p - \Lambda$ where $\Lambda$ could be regarded as the possible time-independent contribution to the energy density and pressure of the vacuum. The detail expressions of $\Lambda$ are already given in original Ref.\cite{LiuWesson}. Then, in order to check the influence of vacuum on the energy flow across horizon, let us look at the total matter (or liquid) components containing both vacuum contribution and the normal matter. We set the total density $\rho_{total} = \rho + \rho_{vac}$ and the total pressure $p_{total} = p + p_{vac}$ where $\rho_{vac}$ and $p_{vac}$ belong to vacuum, $\rho$ and $p$ belong to the normal liquid or matter. Then after considering the standard state equation of vacuum $\rho_{vac} = - p_{vac} = M_{pl}^2 \Lambda$, we easily find the sum of $\rho_{vac}$ and $p_{vac}$ is zero. So the sum of $\rho_{total}$ and $p_{total}$ is equal to the sum of $\rho$ and $p$, i.e. $p_{total} + \rho_{total} = p + \rho$.  Although the Eqs. (\ref{amountenergy}) and (\ref{perconeq}) could include the vacuum, the vacuum contribution does not affect the result. The flow of energy across the horizon depends only on the non-vacuum part. In fact, according to the energy-momentum tensor in Eq.(\ref{1energymomentum}), the parameters $p$ and $\rho$ can belong to any perfect liquids including the vacuum provided the equation of state $p = \gamma \rho$ is satisfied. For different liquids, the values of $\gamma$ are different. For the cosmological constant case (or vacuum case), the state parameter $\gamma$ equals $-1$. Because there is a relationship for vacuum $p_{vac} + \rho_{vac} = 0$, it is also easy to see the vacuum contributes nothing to the flow of energy across the cosmological apparent horizon.
\section{the universal spatially flat model}
It is generally known that the results of recent CMB's astronomical observations incline our universe towards a spatially flat case. In fact if the universe is not flat the geodesics of massless particles such as photons starting out parallel to each other will slowly diverge later \cite{Dodelson}. Meanwhile many other evidences also support the flatness fact, such as the researching of anisotropy spectrum based on three small scale experiments: DASI \cite{Halverson}, Maxima \cite{Lee}, Boomerang \cite{Netterfield}. In the anisotropy spectrum, the pattern of peaks and troughs is shifted to scales smaller than that predicted by open cosmology, clearly favoring the flat case. Moreover the early astrophysical data such as the age of the universe \cite{Leonard,Overduin} also indicates that the spatial sections of FRW universe probably is flat. Just as what is said in Ref.\cite{LiuWesson}, this 5D big bounce scenario can describe many cosmological models by choosing different proper scalar factor functions, especially in the universal spatially flat case. Under the situation we consider, the scale factors of this big bounce are given by \cite{LiuWesson}
\begin{eqnarray}
  \eta(t,y)^2 &=& \mu^2 \left(y - C t\right)^2 + \frac{K}{\mu^2},\label{addeta3dflat}\\
  \zeta(t,y) &=& \frac{\dot{\eta}}{\mu},\label{addzeta3dflat}\\
  \mu(t) &=& t^n, n = -\frac{1 + 3\gamma}{3(1 + \gamma)},\label{addmuandn3dflat}
\end{eqnarray}
where the equation of state is taken to be the isothermal one i.e. $p = \gamma \rho$. The ordinary matter is represented by the state with $\gamma$ in the range $(0, 1/3)$. Two extreme cases which are the dust and radiation and ultrarelativistic particles are represented by the sates of $\gamma = 0$ and $\gamma = 1/3$ respectively. The constant $C$ in the above formula depends on the state parameter $\gamma$,
\begin{equation}\label{square-well}
    C =\left\{
\begin{array}{c}
\frac{3}{2} \left(1 + \gamma\right), \ \ \ 0\leq \gamma<\frac{1}{3},\\
\sqrt{4 - K}, \ \ \ \ \ \ \ \gamma = \frac{1}{3}.\\
\end{array}
\right.
\end{equation}
The density of matter and the cosmological term are listed by
\begin{eqnarray}
  \rho &=& \left(\frac{2}{1 + \gamma}\right) \frac{\mu}{\eta}\left(\frac{\mu}{\eta} - \frac{\dot{\mu}}{\dot{\eta}}\right), \label{2rho}\\
  \Lambda &=& \left(\frac{2}{1 + \gamma}\right) \frac{\mu}{\eta}\left[\left(\frac{1 + 3\gamma}{2}\right)\frac{\mu}{\eta} + \frac{\dot{\mu}}{\dot{\eta}}\right]. \label{2Lambda}
\end{eqnarray}
The apparent horizon is given by
\begin{equation}\label{2apphori}
\widetilde{r}_A = \zeta\frac{\eta}{\dot{\eta}} = \frac{1}{H} = \frac{\sqrt{t^{4n}(y - C t)^2 + K}}{t^{2n}}.
\end{equation}

It is shown that the apparent horizon identically equals the Hubble horizon which is also justified by Ref\cite{cairgKimsp}. Thus, the derivative relationship between the Hubble constant and apparent horizon is given by $\dot{\widetilde{r}}_A = -H \dot{H} \widetilde{r}_A^3$. Then just as what has been done in the former section we assume that the entropy of the apparent horizon takes the normal form $S = A/4G$ and the temperature as $T = 1/2\pi \widetilde{r}_A = H/2\pi$. Hence, according to the amount of energy crossing the apparent horizon, i.e. Eq.(\ref{amountenergy}) and the first law of thermodynamics, the Friedmann equation is also given by
\begin{equation}\label{2feauqtion1}
4\pi G(\rho + p) = -\frac{1}{\zeta} \dot{H}.
\end{equation}
Substituting the continuity equation (\ref{perconeq}) into Eq.(\ref{2feauqtion1}) and integrating the equation, we can obtain another component of Friedmann equations in this spatially flat model,
\begin{equation}\label{2feauqtion2}
\frac{8}{3}\pi G\rho= H^2.
\end{equation}
According to the Hubble constant $H$ in Eq.(\ref{2apphori}) and the spatial Friedmann equation (\ref{2feauqtion2}), the density of big bounce universe with general 3D flat is given by
\begin{equation}\label{2densitybigbounce}
\rho = \frac{3 t^{4n}}{8\pi G \left[t^{4n}\left(y - C t\right)^2 + K\right]}.
\end{equation}
In order to relate the foregoing properties to the subject of apparent horizons, in the following discussion we will consider the original models discussed in Ref. \cite{LiuWesson}, relevant to the cold and hot phases of the universe.
\subsection{the late stage of cold 3D flat universe}
In this subsection, we will discuss the cold 3D flat case. The contribution of the energy component comes chiefly from its rest part rather than its thermodynamic kinetic part. The work of pressure changes only the trivial thermodynamic kinetic energy, and hence the modification of the total mass of $\rho_m R^3$ could be ignored safely. In the natural unit, the pressure of dust is equal to its density of thermal kinetic energy which satisfies $p_{dust} \ll \rho_{dust}$, thus without loss of correctness we can set $\gamma = 0$ and $p_{dust} = 0$ to represent the matter content of the late universe. Hence, in this 5D big bounce universe, the cold 3D flat case is given by
\begin{equation}\label{cold3dflatcase}
\eta^2 = \frac{9}{4} t^{4/3} + K t^{2/3} - 3 y t^{1/3} + y^2 t^{-2/3},
\end{equation}
where $\mu (t) = t^{-1/3}$ and the parameters are listed as $\gamma = 0,\ p = 0,\ n = -1/3,\ C = 3/2$. About the detailed discussion of this case, one can refer to Ref. \cite{LiuWesson}. It's interesting that there is a finite minimum of scalar factor $\eta$ at $t = t_{m1}$, before which our universe contracts and after which it expands. In the usual FRW model, when the time approaches zero the scalar factor vanishes. This situation leads directly to a singularity geometry and the divergent matter, which is called the big bang. However, in this big bounce universe, the 5D curvature invariant in Eq.(\ref{1Kretschmann}) is finite on the hypersurface of $y = \text{constant}$. Furthermore, besides the 5D curvature invariant because $\frac{d \eta}{d t} = 0$ and $\nu (t) \neq 0$ are satisfied at the point of $t_{m1}$, all dynamical quantities in this big bounce universe do not diverge such as $\rho$, $\Lambda$, $^{(4)}R$ and so on.

According to the definition of the Hubble parameter $H(t, y)$ (\ref{1hubble}), we have
\begin{equation}\label{4coldhubble1212}
H^{-1} = \sqrt{\frac{9}{4}t^2 + K t^{4/3} - 3yt + y^2} = \widetilde{r}_A.
\end{equation}

After calculating the amount of energy crossing the apparent horizon, we can get the Friedmann equation in this cold 3D-flat model as
\begin{equation}\label{4coldfeq}
4 \pi G \rho \zeta = - \dot{H}.
\end{equation}
Based on the continuity equation (\ref{perconeq}), the density of this cold case is given by $\rho = 3H^2/8 \pi G$. The integral form of Hubble parameter also could be given by $H^{-1} = \frac{3}{2} \int \zeta (y, t) d t$. Then we consider the late stage of cold 3D flat universe, i.e. $t \gg t_m$, the scalar factor is
\begin{equation}\label{coldAA}
\eta = \frac{3}{2} t^{2/3} \left[1 + \frac{2}{9} K t^{-2/3} - \frac{2}{3} y t^{-1} + \mathcal{O}(t^{-4/3}) \right],
\end{equation}
\begin{equation}\label{coldBB}
\zeta = \frac{\dot{\eta}}{\mu} = 1 + \frac{y}{3t} + \mathcal{O}(t^{-4/3}).
\end{equation}

The 5D line element reads
\begin{equation}\label{coldelement}
d S^2 = \left[1 + \mathcal{O}(t^{-1})\right] d t^2 - \left[\frac{3}{2} t^{2/3} + \frac{1}{3}K + \mathcal{O}(t^{-1/3})\right]^2\left(d r^2 + r^2 d \Omega^2\right) - d y^2,
\end{equation}
which is an Einstein de Sitter like spacetime. When $t$ approaches proper time, the scale factor varies as $t^{2/3}$.

According to the definition of the Hubble parameter $H(t, y)$ (\ref{1hubble}), we have
\begin{equation}\label{4coldhubble++}
H^{-1}= \frac{2}{3} \left(t + \frac{1}{3} y \ln t\right) + C_1,
\end{equation}
where $C_1$ is an integral constant.
Hence, the density of this case is shown by
\begin{equation}\label{4densityhubble}
\rho = \frac{243}{8\pi G \left(6 t + 2 y \ln t + 9 C_1\right)^2}.
\end{equation}
\begin{figure}
  \includegraphics[width=4 in]{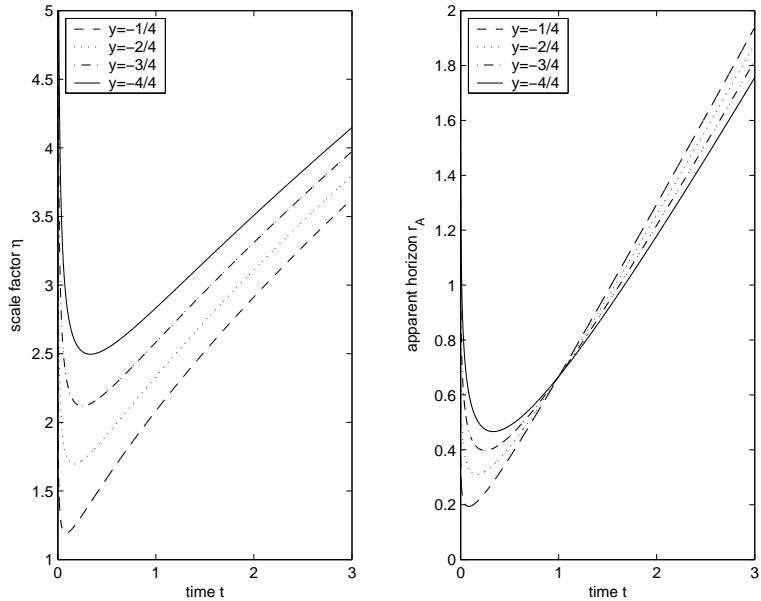}\\
  \caption{the apparent horizons vs time in the $t\gg t_m$ case in the cold 3D flat model with $K = 1$ and $y = -1/4$(dashed line), $-2/4$(dotted line), $-3/4$(dash-dot line), $-1$(solid line).}\label{fig11}
\end{figure}

The apparent horizons of cold 3D flat model are illustrated in Fig \ref{fig11} where the big bounce's scale factor shown first by Ref.\cite{LiuWesson} are also listed to facilitate comparison. The apparent horizons coincide with each other at the time $t = 1$ where intersection point has nothing to do with the extra dimension which could be seen easily by
the definition of the apparent horzion in Eq.(\ref{4coldhubble++}). Before the time of intersection, $\widetilde{r}_A$ decreases with bigger extra dimension and after that $\widetilde{r}_A$ increases with bigger extra dimension. Similar with the scalar factor's bounce point $t_{m1}$, there is also a finite minimum of $\widetilde{r}_A$ at $t = t_{m2}$, before $t_{m2}$ the apparent horizon contracts and after $t_{m2}$ it expands. In order to compare with universe's big bounce, it is necessary to have a look at the difference between $t_{m1}$ and $t_{m2}$ under the late time condition. After differentiating Eq.(\ref{1hubble}) with respect to time in the cold case, we can get a formulas $\dot{\widetilde{r}_A} = \dot{\eta} t^{1/3} + \eta/3 t^{2/3}$ where the last term is negligible for late universe. Hence, one can find that with a given extra dimension the bounce time of apparent horizon $t_{m2}$ has the same magnitude with universe's big bounce time $t_{m1}$ in late universe algebraically. The result indicates that in the late stage of cold 3D flat universe the apparent horizon contracts before bounce time and expands after it, in agreement with the evolution of the big bounce's scale factor $\eta$ exactly.
\subsection{the late stage of hot 3D flat universe}
In this part we will discuss the late universe dominated by the radiation or ultrarelativistic particles, such as the photon radiation. Because the rest mass of photon is null and the thermal kinematic velocity is the speed of light, the whole density of photon comes from its kinematic mass completely. Based on the Planck statistical distribution of photon, the equation of state is $P_{\gamma} = \rho_{\gamma}/3$. Hence, in this 5D big bounce universe, the hot 3D flat case is given by
\begin{equation}\label{5hota}
\eta^2 = 4 t - 2 C y + \frac{y^2}{t},
\end{equation}
where $\gamma = 1/3$, $n = -1/2$, $C = \sqrt{4 - K}$ and $\mu = t^{-1/2}$. The behavior of this hot universe is similar with the cold case, which contracts before the time of big bounce $t_{m1}$ and expands after $t_{m1}$. For further discussions, one can refer to Ref.\cite{LiuWesson}.

The Hubble parameter of this hot 3D flat model is given by
\begin{equation}\label{5hotcaseHubble}
    H^{-1} = \sqrt{4 t^2 - 2 y t \sqrt{4 - K} + y^2}.
\end{equation}
Using the former apparent horizon method, the Friedmann equations of the hot case are given by $16\pi G \rho \zeta = -3\dot{H}$. By using the continuity equation, the Hubble parameter is given by $H^{-1} = 2 \int \eta (y, t) d t$. When $t \gg t_m$ the first term of $\eta^2$ in Eq.(\ref{5hota}) is dominant, we can have the following approximate expressions
\begin{eqnarray}
\eta &=& 2 t^{1/2} \left(1 - \frac{\sqrt{4-K}}{4t} y + \mathcal{O} (t^{-2})\right),\label{5hot1aa} \\
\zeta &=& 1 + \frac{y}{4t}\sqrt{4 - K} + \mathcal{O} (t^{-2}).
\end{eqnarray}

The 5D radiation metric is obtained as
\begin{equation}\label{5hot1metric}
d S^2 = \left[1 + \frac{Cy}{2t} + \mathcal{O} (t^{-2})\right] dt^2 - 4t \left[1 - \frac{Cy}{2t} + \mathcal{O}(t^{-2})\right]\left(d r^2 + r^2 d \Omega^2\right) - d y^2,
\end{equation}
which is analogous to the 4D radiation metric. Unlike the cold case, when $t$ approaches proper time, the scale factor varies as $t^{1/2}$.

The Hubble parameter is
\begin{equation}\label{5hothubble}
H^{-1} = 2t + \frac{y}{2} \sqrt{4 - K}\ln t + C_3
\end{equation}
where $C_3$ is the integral constant. The energy density in this case is shown by
\begin{equation}\label{5hot1density}
\rho = \frac{3}{2\pi G \left(4 t + y \sqrt{4 - K} \ln t + 2 C_3\right)^2}.
\end{equation}
\begin{figure}
  \includegraphics[width=4 in]{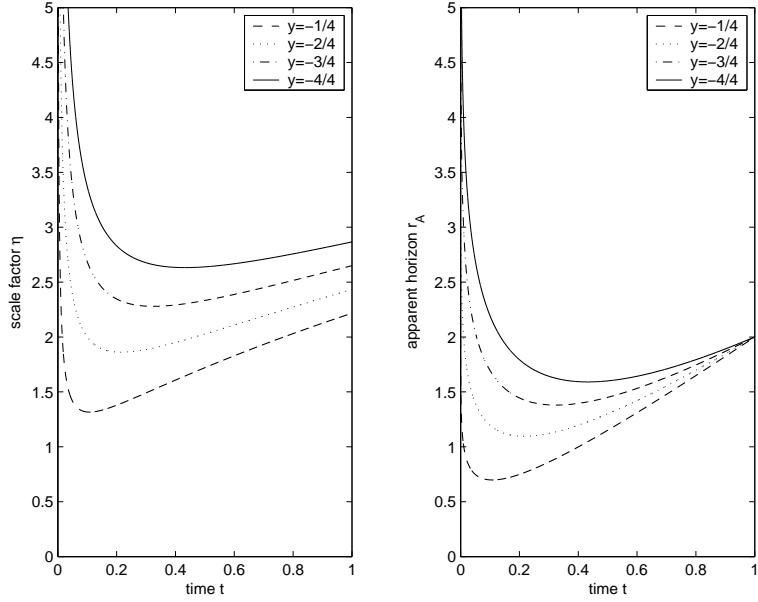}\\
  \caption{the apparent horizons vs time in the $t\gg t_m$ case in the hot 3D flat model with $K = 1$, $C_3 = 0$ and $y = -1/4$(dashed line), $-2/4$(dotted line), $-3/4$(dash-dot line), $-1$(solid line).}\label{fig33}
\end{figure}

It is illustrated in Fig.\ref{fig33} that in this late hot case the apparent horizons take the same value at the point of $t = 1$ which again has nothing to do with the extra dimension. With larger extra dimension $y$, the apparent horizons decrease accordingly. According to Eq.(\ref{1hubble}), we can get $\dot{\widetilde{r}_A} = \dot{\eta} t^{1/2} + \eta/2 t^{1/2}$ where the last term can be neglected for late time. Hence, like the former cold case, the bounce point $t_{m2}$ of apparent horizons also has the same magnitude with the bounce point $t_{m1}$ of the scale factor.
\subsection{the early stage of cold and hot 3D flat universe}
For the early stage of the cold 3D flat universe, the last term in $\eta^2$ (\ref{cold3dflatcase}) is dominant with $t \ll t_m$. So we have the approximate expressions of $\eta$ and $\zeta$ shown by
\begin{eqnarray}
\eta &=& y t^{-1/3} \left(1 - \frac{3}{2y} t + \frac{K}{2y^2} t^{4/3} + \mathcal{O} (t^2)\right), \label{4coldaa}\\
\zeta &=& -\frac{y}{3t} - 1 + \frac{K}{2y}t^{1/3} + \mathcal{O} (t^2), \label{4coldbb}
\end{eqnarray}
The 5D line element has the canonical form as
\begin{equation}\label{4coldds2}
d S^2 = \frac{y^2}{L^2}\left[d T^2 - e^{-2T/L} \left(d r^2 + r^2 d \Omega^2\right) \right] - d y^2,
\end{equation}
where the coordinate transformation $t = L^3 e^{3T/L}$ is adopted. This metric is of a canonical form given by Ref.\cite{LiuMashhoon}.

The Hubble parameter is
\begin{equation}\label{4coldhubble}
H^{-1} = \frac{y}{2} \ln \frac{1}{t} - \frac{3}{2} t + \frac{9}{16} \frac{K}{y} t^{4/3} + C_2,
\end{equation}
where $C_2$ is the integral constant. Hence the energy density of this case is shown by
\begin{equation}\label{4hotdensity}
\rho = \frac{96}{\pi G \left(8 y \ln 1/t - 24 t + 9 K/y t^{4/3} + 16 C_2\right)^2}.
\end{equation}

For the early stage of the hot 3D flat universe, the last term of $\eta^2$ (\ref{5hota}) is dominant with $t \ll t_m$, which gives us the following formulas
\begin{eqnarray}
  \eta &=& y t^{-1/2} \left(1 - \frac{\sqrt{4 - K}}{y} t + \mathcal{O}(t^2)\right), \label{5hot2eta1} \\
  \zeta &=& -\frac{y}{2t} - \sqrt{1 - \frac{K}{4}} + \mathcal{O}(t^2), \label{5hot2zeta1}
\end{eqnarray}
The canonical form of metric could be obtained as
\begin{equation}\label{5hot2metric}
d S^2 = \frac{y^2}{\mathcal{L}^2}\left[d T^2 - e^{-2T/L}\left(dr^2 + r^2 d\Omega^2\right)\right] - d y^2,
\end{equation}
where the coordinate transformation $t = \mathcal{L}^2 e^{2T/\mathcal{L}}$ is used. Comparing with the cold case, one can easily find that this canonical metric solution (\ref{5hot2metric}) is the same as the former case in Eq.(\ref{4coldds2}).

The Hubble parameter is
\begin{equation}\label{5hot2hubble}
H^{-1} = y \ln \frac{1}{t} - \sqrt{4 - K} t + C_4,
\end{equation}
where $C_4$ is the integral constant. The energy density of this case is shown by
\begin{equation}\label{5hot2density}
\rho = \frac{3}{8 \pi G \left(y \ln 1/t - \sqrt{4 - K} t + C_4\right)^2}.
\end{equation}
It looks a little surprising that both in the cold case (\ref{4coldaa}) and in the hot case (\ref{5hot2eta1}), the geometries have the same form exactly. In order to see why it is this, we need to start from the very beginning, i.e. the universal formulas Eqs.(\ref{addeta3dflat}), (\ref{addzeta3dflat}) and (\ref{addmuandn3dflat}) describing the 3D flat model no matter in cold case or in hot case. Then considering the situation of the early universe stage $t \ll t_{m1}$, the scalar factor is rewritten approximately as
\begin{equation}\label{6eta}
\eta(t, y)^2 = y^2 t^{2n} + K t^{-2n}.
\end{equation}
Calculating the partial derivative of $\eta(t, y)^2$ with respect to $t$ directly, one can get the expression of $\zeta (t, y)$
\begin{equation}\label{6etaeta}
\frac{\dot{\eta}}{\mu} = \left(y^2 - \mu^{-4} K\right) \frac{\dot{\mu}}{\eta^2} = \zeta(t, y),
\end{equation}
which gives us $\zeta(t, y)^2$ appeared in the metric,
\begin{equation}\label{6zata2}
\zeta(t, y)^2 = \frac{n^2}{t^2} \frac{\left(y^2 - K t^{-4n}\right)^2}{y^2 + K t^{-4n}}.
\end{equation}

Then in this big bounce universe, the parameter $n$ in Eq.(\ref{addmuandn3dflat}) is negative if the state parameter $\gamma$ is positive. So the negative $n$ ensures the term $t^{-2n} K$ contained in Eqs.(\ref{6eta}) and (\ref{6zata2}) could be safely ignored in the early stage of the 3D flat universe. Finally, the expressions of $\eta^2$ and $\zeta^2$ can be rewritten in the simple formulas as
\begin{eqnarray}
  \eta(t, y)^2 &=& y^2 t^{2n}, \label{66eta} \\
  \zeta(t, y)^2 &=& y^2 \frac{n^2}{t^2}, \label{66zata}
\end{eqnarray}
Then we carry out the coordinate transformation $t = L^{-1/n} e^{-\tau/nL}$ where $L$ is a constant, the big bounce metric in Eq.(\ref{15dmetric}) becomes
\begin{equation}\label{66dmetric}
d S^2 = \frac{y^2}{L^2}\left[d \tau^2 - e^{-2\tau/L}\left(d r^2 + r^2 d \Omega^2\right)\right] - d y^2,
\end{equation}
which is a canonical form of metric shown by Refs.\cite{LiuMashhoon,wessonstm}. The brackets of metric in Eq.(\ref{66dmetric}) is the 4D de Sitter metric which could be interpreted as $\rho = 0$ and $\Lambda = 3/L^3$. Substituting Eq.(\ref{66eta}) into the density in Eq. (\ref{2rho}) and the cosmological term in Eq.(\ref{2Lambda}), with general state parameter $\gamma$ we can get $\rho = 0$ and $\Lambda = 3/y^2$ which is justified as the limit of $t \rightarrow 0$ in the early stage of cold and hot 3D flat universe shown by Ref.\cite{LiuWesson}. Moreover, according to Eq.(\ref{66eta}) the apparent horizon satisfies $\widetilde{r}_A = y$ which shows again that the extra dimension affects the early universe more deeply. Then according to the coordinate transformation $t = L^{-1/n} e^{-\tau/nL}$, one can see that $t \longrightarrow 0$ corresponds to $\tau \longrightarrow - \infty$, in which $\tau$ is the proper time of 4D. So the universe exists forever before the big bounce in $\tau$ time. So in the early stage of the 3D flat universe, the geometries of spacetime, which contain the 4D de Sitter metric, have the same forms exactly no matter for the cold case or the hot case. This result does not depend on the equation of state and has nothing to do with the specific form of matter.

It should be noted that the same line elements containing a 4D de Sitter metric at the early stage for both cold and hot models mean that they are equivalent to the vacuum metric in the early stages their evolution. However, this interesting result cannot be applied to late stages. It is because the proper time $\tau$ of the two cases are different in that the coordinate transformation $t = L^{-1/n} e^{-\tau/nL}$ depends on the choice of parameter $n$ shown in Eq.(\ref{addmuandn3dflat}). For the cold model with $\gamma = 0$ the relationship between $t$-time and $\tau$-time is $t = L^3 e^{3\tau/L}$ with $n = -1/3$, and for the hot model with $\gamma = 1/3$ the relationship between $t$-time and $\tau$-time is $t = L^2 e^{2\tau/L}$ with $n = -1/2$. These two kinds coordinate transformations are first adopted in the original literature \cite{LiuWesson}, in this paper we unify them together as one form by the field parameter $n$. With various choices of $n$, the values of the density $\rho$, scale factor $\eta$, cosmological constant $\Lambda$, etc. have different magnitudes. The different densities $\rho$ are shown by Eqs.(\ref{4hotdensity}) and (\ref{5hot2density}) through the flat Friedmann equation $8 \pi G \rho/3 = H^2$. Various scale factors $\eta$ are given by Eqs.(\ref{4coldaa}) and (\ref{5hot2eta1}), and the cosmological constant $\Lambda$ given by Eq.(\ref{2Lambda}). The detailed expressions of $\Lambda$ for these two extreme cases are given in Ref.\cite{LiuWesson}.
\section{conclusion}
In this paper, we study the thermodynamic properties of the apparent horizon in the 5D big bounce universe. We summarize what has been achieved. Firstly, the connections between the thermodynamics and gravity are obtained in big bounce universe. The Friedmann equations of universe can be obtained by using the first law of thermodynamics, and the perfect fluid's continuity equation (\ref{perconeq}) to the apparent horizon with general spatial curvature. Meanwhile, we also check the result in two extreme cases under the spatially flat universe, one is the dust case described by the matter content of the late universe with $\gamma = 0$ and $p = 0$, another is the case of radiation or ultrarelativistic particles described the early universe with $\gamma = 1/3$ and $p = \rho/3$. The continuity equation used here is different from that of the normal 4D case, i.e. $\dot{\rho} + 3 H (\rho + p) = 0$. The reason is that Hubble parameters Eq.(\ref{1hubble}) of big bounce universe are determined both by the scale factor of 3D space $\eta$ (\ref{addeta}) and the scale factor of the time $\zeta$ (\ref{addzeta}). This situation is unlike the Hubble parameter of the normal 4D universe which is determined only by the scale factor of 3D space.

Secondly, the bounce points illustrated in Figs.\ref{fig11} and \ref{fig33} also exist in the apparent horizon for the late cold and hot universes. In these two cases, the apparent horizons decrease monotonously before bounce points and increase monotonously after them which is in agreement with universe's scalar factor $\eta (y, t)$. It is also found that in these late cases the apparent horizons cross at the point of $t = 1$ which are unaffected by the extra dimension. With larger extra dimension, the apparent horizon decreases before $t = 1$ and increases after this point. In the early stage of universe, it is interesting that based on a general state parameter $\gamma$ the original big bounce metric (\ref{15dmetric}) reduces to a canonical form metric (\ref{66dmetric}) which contains a 4D de Sitter metric. This result is independent on the equation of state. The apparent horizon of the early universe, i.e. $\widetilde{r}_A = y$, is dependent on the extra dimension which indicates the extra dimension affects the early universe strongly.

Thirdly, the surface gravity $\kappa$ is a classical quantity which plays an important role in black hole's thermodynamics. However, the generally accepted definition of $\kappa$ only exists near Killing horizons in stationary space-times \cite{Wald,Hawking}, which is based on the existence of a global time translational Killing vector field that becomes null on the event horizon. In a dynamical setting the Killing vector doesn't exist
so the Killing definition of surface gravity is no longer valid. Various dynamical
definitions are proposed to address this problem \cite{Collins,Fodor,Abreu,Visser,Nielsen1,Nielsen2,Hayward1}. In this paper, we adopt Bak and Rey's method to define the surface gravity $\kappa$ associated with the dynamical apparent horizon, i.e. Eq.(\ref{addsurface}), which could be treated as a generalization of Kodama vector definition proposed by Hayward et al. \cite{Hayward1}. According to this kind definition of surface gravity in Eq.(\ref{addsurface}), we can get a direct calculation given by
\begin{equation}\label{conclusionkappa}
\kappa = -\frac{1}{\widetilde{r}_A} \left(1 - \frac{\dot{\widetilde{r}_A}}{2 H \widetilde{r}_A}\right).
\end{equation}
Then based on the first law of thermodynamics $-d E = T d S$ which is assumed to hold for the dynamical apparent horizon, the apparent horizon radius $\widetilde{r}_A$ should be regarded to have a fixed magnitude after considering an infinitesimal amount of energy crossing the apparent horizon. So in this way, we have a approximate relationship $\kappa \approx -1/\widetilde{r}_A$ which correctly recover $T \equiv |\kappa|/2\pi = 1/2\pi\widetilde{r}_A$ at the apparent horizon. All in all the horizon area and surface gravity are purely geometric quantities given by the spacetime geometry, and the entropy and temperature have a certain universality with the horizon.

Fourthly, we also would like give some words about the relationship between the Hubble constant and the entropy associated with the apparent horizon. According to the expression of apparent horizon in Eq.(\ref{1apphor1}) derived from the vector $\nabla \widetilde{r}$ in Eq.(\ref{addapparenthorizon}), one can see that if the spatial curvature vanishes, the apparent horizon is equal to the reciprocal of the Hubble constant, which is already justified by many literatures \cite{cairgKimsp,Bak}. Then the thermodynamics is applied to the apparent horizon. In this way, we can construct the relationship between the Hubble constant and the entropy of the apparent horizon.

Finally, it should be mentioned that this cosmological model is compatible with the observational consequences under the appropriate choice of parameters. About this aspect there are many literatures \cite{addobsers1,addobsers2,addobsers3,addobsers4,addobsers5,addobsers6}. Three components contained in the matter, which are CDM+baryons $\rho_m$, radiation $\rho_r$ and dark energy $\rho_x$, are first studied in \cite{addobsers1}. These three components are assumed to be minimally coupled with each others and the corresponding pressures are given by $p_m = 0$, $p_r = \rho_r/3$ and $p_x = w_x \rho_x$. The dimensionless density parameters for CDM+baryons, radiation and dark energy are given by $\Omega_m = \rho_m/(\rho_m + \rho_r + \rho_x)$, $\Omega_r = \rho_r/(\rho_m + \rho_r + \rho_x)$ and $\Omega_x = 1 - \Omega_m -\Omega_r$. For the simplest case, $\mu (t) = a t + b/t$ and $\nu(t) = c t$ are assumed. In order to match the observational data, the parameters are chosen as followings: $K = 1,\ \ y=1,\ \ \rho_{m0} = 1.1,\ \ \rho_{r0} = 2.4, a = 9 \times 10^{-6}, b = 3.5, c = 0.11$.
One of results shows that the proportion of CDM+baryons and dark energy are $\Omega_m \sim 0.3$ and $\Omega_x \sim 0.7$, respectively. Meanwhile, the results give the redshift $z_T \approx 0.37$ for the transition from decelerated expansion to accelerated expansion which agrees with $z_T = 0.46 \pm 0.13$ obtained from the observations through SN Ia luminosity distance. The equation of state of dark energy $w_{x0} \approx -1.05$ is obtained as well. Later, the metric functions $\eta$ and $\zeta$ are converted into a new function $f(z)$ with the cosmological red shifts as the only variable \cite{addobsers2,addobsers4}. Function $f(z)$ could be determined according to the universe's parameterization. In Ref.\cite{addobsers2}, two components of CDM+baryons and dark energy are considered in which the function $f(z)$ takes on an equivalent function as the potential $V(\phi)$ in quintessence or phantom models.
Their results show that a scaling stage exists before $z \sim 2$ where the ratio of dark
energy to CDM+baryons could be constant and the coincidence problem could be avoided.
At $z_T \sim 0.8$ there is a transition from decelerated expansion to accelerated expansion.
Other predictions including densities $\Omega_{mzE}$ and $\Omega_{m0}$, the deceleration parameters $q_0$ and so on in this big bounce universe are consistent with the current observations. Until now, we still do not know too much about the detail of the universe including the dark energy, dark matter and so on. The complete parameterization beyond the models are not finished. However, many results obtained in this model are compatible with the observations. About the detailed content about the parameter chosen to fit the observational data on the real universe, one can refer to the literatures \cite{addobsers1,addobsers2,addobsers3,addobsers4,addobsers5,addobsers6}.

\acknowledgments
We are grateful to the anonymous referees for insightful comments, and to Y. Han for helpful reading. Liu's work is supported in part by NSFC (No.11475143), HASTIT (No.14HASTIT043) and Foundation for University Key Teacher by He'nan Educational Committee. Xu's work is supported in part by NSFC(No.11275035). Lv's work is supported in part by NSFC(No.11205078).


\begin{thebibliography}{*}
\bibitem{Maldacena}J.M. Maldacena, Adv. Theor. Math. Phys. 2 (1998) 231 [arXiv:hep-th/9711200].
\bibitem{Jacobson}T. Jacobson, Phys. Rev. Lett. 75 (1995) 1260 [arXiv:gr-qc/9504004].
\bibitem{Brustein}R. Brustein and M. Hadad, Phys. Rev. Lett. 103 (2009) 101301; Erratum-ibid. 105 (2010) 239902 [arXiv:0903.0823].
\bibitem{Padmanabhan}T. Padmanabhan, [arXiv:0903.1254].
\bibitem{Parikh}M. K. Parikh and S. Sarkar, [arXiv:0903.1176].
\bibitem{Padmanabhan2}T. Padmanabhan, Class. Quant. Grav. 19 (2002) 5387 [arXiv:gr-qc/0204019].
\bibitem{cairgKimsp}R.G. Cai and S.P. Kim, JHEP 0502 (2005) 050 [arXiv:hep-th/0501055].
\bibitem{frgravity}C. Eling, R. Guedens and T. Jacobson, Phys. Rev. Lett. 96 (2006) 121301
[arXiv:gr-qc/0602001]; M. Akbar and R.G. Cai, Phys. Lett. B 648 (2007) 243 [arXiv:gr-qc/0612089].
\bibitem{scalartensor}M. Akbar and R. G. Cai, Phys. Lett. B 635 (2006) 7 [arXiv:hep-th/0602156]; R.G. Cai and L.M. Cao, Phys. Rev. D 75(2007) 064008 [arXiv:gr-qc/0611071].
\bibitem{brancase}R. G. Cai and L. M. Cao, Nucl. Phys. B 785 (2007) 135-148 [arXiv:hep-th/0612144]; A. Sheykhi, B. Wang and R. G. Cai, Nucl. Phys. B 779 (2007) 1-12 [arXiv:hep-th/0701198]; A. Sheykhi, B. Wang, R.G. Cai, Phys. Rev. D76 (2007) 023515 [arXiv:hep-th/0701261].
\bibitem{Padmanabhan4}T. Padmanabhan, Phys. Rept. 406 (2005) 49-125 [arXiv:gr-qc/0311036].
\bibitem{Padmanabhan3}T. Padmanabhan, [arXiv:gr-qc/0910.0839].
\bibitem{Steinhardt}P. J. Steinhardt and N. Turok, Science, 296 (2002) 1436 [arXiv:hep-th/0111030].
\bibitem{Tolman}R. C. Tolman, Relativity, Thermodynamics and Cosmology, (Oxford U. Press, Clarendon Press,1934).
\bibitem{Khoury1}J. Khoury, B. A. Ovrut, P. J. Steinhardt and N. Turok, Phys. Rev. D 64 (2001) 123522 [arXiv:hep-th/0103239].
\bibitem{Khoury2}J. Khoury, B.A. Ovrut, N. Seiberg, P.J. Steinhardt and N. Turok, Phys. Rev. D 65 (2002) 086007 [arXiv:hep-th/0108187].
\bibitem{LiuWesson}H. Liu and P.S. Wesson, Astrophys. J. 562 (2001) 1-6 [arXiv:gr-qc/0107093].
\bibitem{LiuMashhoon}H. Liu and B. Mashhoon, 1995, Ann. Phys. (Leipzig) 4, 565.
\bibitem{Robertson}H.P. Robertson and T.W. Noonan, 1968, Relativity and Cosmology (Philadelphia: Saunders).
\bibitem{Bak}D.S. Bak and S.J. Rey, Class. Quant. Grav. 17 (2000) L83, [arXiv:hep-th/9902173].
\bibitem{Wald}R.M. Wald, General Relativity (Univ. Chicago Press, Chicago, 1984).
\bibitem{Hawking}S.W. Hawking and G.F.R. Ellis, The Large Scale Structure of Space-Time (Univ. Cambridge Press, Cambridge, 1973).
\bibitem{Dodelson}S. Dodelson, Modern Cosmology (Elsevier, 2008).
\bibitem{Halverson}N.W. Halverson et al., Astrophysical Journal 568 (2002) L28.
\bibitem{Lee}A.T. Lee et al., Astrophysical Journal 561 (2001) L1.
\bibitem{Netterfield}C.B. Netterfield et al., Astrophysical Journal 571 (2002) 604.
\bibitem{Leonard}S. Leonard and K. Lake, Astrophysical Journal 441 (1995) 548.
\bibitem{Overduin}J. M. Overduin, Astrophysical Journal 517 (1999) L1.
\bibitem{wessonstm}P.S. Wesson, Space-Time-Matter (Singapore: World Scientific, 1999).
\bibitem{Nielsen1} A. B. Nielsen and J. H. Yoon, Class. Quant. Grav. 25 (2007) 085010.
\bibitem{Collins} W. Collins, Phys. Rev. D 45 (1992) 495.
\bibitem{Nielsen2} A. B. Nielsen and M. Visser, Class. Quant. Grav. 23 (2006) 4637.
\bibitem{Hayward1} S. A. Hayward, R. Di Criscienzo, M. Nadalini, L. Vanzo and S. Zerbini, Class. Quant. Grav. 26 (2009) 062001.
\bibitem{Fodor} G. Fodor et al., Phys. Rev. D 54 (1996) 3882.
\bibitem{Visser} M. Visser, Int. J. Mod. Phys. D 12 (2003) 649.
\bibitem{Abreu} G. Abreu and M. Visser, Phys. Rev. D 82 (2010) 044027.
\bibitem{addobsers1}L.X. Xu and H. Y. Liu, Int. J. Mod. Phys. D14 (2005) 883-892, [arXiv:astro-ph/0412241].
\bibitem{addobsers2}L.X. Xu and H.Y. Liu, Int. J. Mod. Phys. D14 (2005) 1947-1958, [arXiv:astro-ph/0507250].
\bibitem{addobsers3}L.X. Xu, H.Y. Liu and B.R. Chang, Mod. Phys. Lett. A20 (2005) 3105-3114, [arXiv:astro-ph/0507397].
\bibitem{addobsers4}L.X. Xu, H.Y. Liu and C. W. Zhang, Int. J. Mod. Phys. D15 (2006) 215-224, [arXiv:astro-ph/0510673].
\bibitem{addobsers5}C.W. Zhang, H.Y. Liu, L.X. Xu and P. S. Wesson, Mod. Phys. Lett. A21 (2006) 571-580, [arXiv:astro-ph/0602414].
\bibitem{addobsers6}B.R. Chang, H.Y. Liu, L.X. Xu, C.W. Zhang, Mod. Phys. Lett. A23 (2008) 269-279, [arXiv:astro-ph/0704.3670].
\end{thebibliography}
\end{document}